\documentstyle[12pt,preprint,aps,epsf]{revtex}
\newcommand{\be}{\begin{equation}}
\newcommand{\ee}{\end{equation}}
\newcommand{\bea}{\begin{eqnarray}}
\newcommand{\eea}{\end{eqnarray}}
\newcommand{\bdm}{\begin{displaymath}}
\newcommand{\edm}{\end{displaymath}}
\begin{document}
\large{

\title{Hysteretic Response of an Anti-Ferromagnetic Random-Field 
Ising Model in One Dimension at Zero Temperature}

\author{  Prabodh Shukla, Ratnadeep Roy, and Emilia Ray
\\Physics Department \\ North Eastern Hill University
\\ Shillong-793 022, INDIA }
\maketitle

\begin{abstract}
We consider the hysteretic response of a
one-dimensional anti-ferromagnetic random-field Ising model at
zero temperature for a uniform bounded distribution of quenched
random fields, and present analytic results in a limited range
of the applied field.
\end{abstract}

\vspace{5cm}
email: shukla@dte.vsnl.net.in
\clearpage
\section{Introduction}

Relaxation dynamics of random field Ising model (RFIM) at zero
temperature provides a simple caricature of complex non-equilibrium
phenomena. RFIM with ferromagnetic as well as anti-ferromagnetic
interactions has been investigated in the context of two distinct
classes of relaxational behavior. The ferromagnetic RFIM shows
relaxation by avalanches and has been applied ~\cite{sethna,dhar}
to study hysteresis and Barkhausen jumps in magnetic materials.
The anti-ferromagnetic   model does not support avalanches, and
is more appropriate ~\cite{shukla1,kisker} for the glassy
kind of dynamics where relaxation may proceed by single
localized events. In spite of the simplicity of RFIM
and its relaxational dynamics at zero temperature, exact solution
of the model is difficult on account of its randomness. So far,
analytic solution of the model has been obtained in one
dimension~\cite{shukla2} and on a Bethe lattice~\cite{dhar,sanjib}
for ferromagnetic interactions only. In this paper, we present
a solution of the non-equilibrium dynamics of the
anti-ferromagnetic RFIM at zero temperature in one dimension in
a limited range of the applied field. For simplicity, we limit
ourselves to a bounded uniform distribution of the random field.
It is possible to extend this solution to other distributions of
the random field, but this will not be taken up in this paper.

\section{The Model}
At each site of a one dimensional lattice, there is an Ising spin
$ s_{i}=\pm{1}$, $i=1,2,3......,n$ which interacts with its nearest
neighbors through an anti-ferromagnetic interaction J. Each site
has a quenched random field $h_{i}$ drawn independently from a
continuous bounded distribution,

\begin{eqnarray}
p(h_{i}) & = \frac{1}{2\Delta} &\qquad\mbox{if}\qquad
-\Delta \le h_{i} \le \Delta
\nonumber \\ & = 0, & \qquad\mbox{    otherwise.}
\end{eqnarray}

The entire system is placed in an externally applied uniform field $h_{a}$.
The Hamiltonian of the system is given by
\be
H=-J \sum_{i}{s_{i}s_{i+1}}-\sum_{i}h_{i}s_{i}-h_{a}\sum s_{i}
\ee
We consider the hysteretic response of this system when the
external field $h_{a}$ is slowly increased  from $-\infty$ to $+\infty$.
We assume the dynamics to be the single-spin-flip Glauber dynamics
at zero temperature, i.e. a spin is flipped only if it lowers
the energy. We assume that if the spin-flip is allowed, it occurs with
a rate $\Gamma$, which is much larger than the rate at which
the magnetic field $h_{a}$ is increased. Thus we assume that all flipable
spins relax instantly, and the spin $s_{i}$ always has the same
sign as the net local field $l_{i}$ at the site.
\be
s_{i}=\mbox{ sign }{l_{i}} =\mbox{ sign }
[J(s_{i-1}+s_{i+1})+h_{i}+h_{a}]
\ee

The hysteretic response of the system to an
applied field $h_{a}$ is measured by
the magnetization $m(h_{a})$ per spin, 
\be
m(h_{a})=\frac{1}{N}\sum_{i}s_{i}
\ee
We start with $h_{a}=-\infty$, when $m=-1$, and increase $h_{a}$
slowly to
$h_{a}=+\infty$ when m=1. Our object is to calculate $m(h_{a})$
for all values of $h_{a}$. The magnetization $m_{R}(h_{a})$ on the
return trajectory when $h_{a}$ is slowly decreased from
$h_{a}=+\infty$
to $h_{a}=-\infty$ can be obtained from $m(h_{a})$ by a
symmetry relation
$m_{R}(h_{a})=-m(-h_{a})$. Therefore the knowledge of the magnetization
curve on the lower half of the hysteresis loop suffices to
determine the entire hysteresis loop, and we can limit ourselves
to the calculation of $m(h_{a})$ alone.

\section{simulations}
It is instructive to look  at a computer simulation of the
model before proceeding to obtain it analytically. Figure 1
shows magnetization $m(h_{a})$ in an increasing applied
field $h_{a}$  for an anti-ferromagnetic RFIM ($J=-1$),
obtained from a simulation of $10^{3}$ spins
averaged over $10^{3}$ different realizations of the random
field distribution of width $2\Delta=1$. The magnetization
$m(h_{a})$ rises from -1 to +1 in three
steps. We call these steps as ramp-I ($h_{a}=-2|J|-\Delta$
to $h_{a}=-2|J|+\Delta$); ramp-II ($h_{a}=-\Delta$
to $h_{a}=+\Delta$); and ramp-III ($h_{a}=2|J|-\Delta$
to $h_{a}=2|J|+\Delta$). The ramps are connected to each other by
two plateaus; plateau-I ( $h_{a}=-2|J|+\Delta$ to $h_{a}=-\Delta$);
and plateau-II ($h_{a}=+\Delta$ to $h_{a}=2|J|-\Delta$). On the
plateaus, the magnetization remains constant even though the
applied field continues to increase. Plateaus occur for
$\Delta \le |J|$ (small disorder), and simulations suggest that
magnetization on the plateaus is independent of $\Delta$.
Numerically, the magnetization on the plateaus is
approximately $m^{I}=-.135$ on plateau-I, and $m^{II}=.109$
on plateau-II.

The qualitative shape of $m(h_{a})$ is easy to understand. Due to
the anti-ferromagnetic interaction between nearest neighbors,
spins with both  neighbors down are the easiest to be turned
up by an applied field increasing in the up direction. Such
spins turn up on ramp-I. Next are the spins with one
neighbor up and one down which turn up on ramp-II. Spins
with both neighbors up require the largest applied field to
turn up, and these turn up on ramp-III. On each ramp, the sequence
in which the spins turn up is determined by the distribution of
the quenched random field. Spins with large positive quenched
field turn up before spins with a lower quenched field. The
quenched field lies in the range $-\Delta$ to $+\Delta$.
Thus each ramp has a width $2\Delta$ along the axis of
the applied field.

When a spin turns up on ramp-I, its nearest neighbors are 
placed  in a category so that they cannot turn up before
ramp-II. Similarly when a spin turns up on ramp-II, its nearest
neighbor which is down cannot turn up before ramp-III. This is
essentially the reason for the absence of avalanches in the
anti-ferromagnetic RFIM. Occasionally on  ramp-II and ramp-III,
a spin turning up can induce its nearest neighbor which is
already up to turn down. We will discuss the situation on
ramp-II in detail later, but suffice it to say here that this
process too does not cause an avalanche. With
anti-ferromagnetic interactions, spins turn up one at a time
(no Barkhausen noise), and the calculation of $m(h_{a})$ becomes
essentially a matter of sorting quenched random fields in
decreasing order on each ramp. The difficulty arises from
the fact that the a posteriori distribution of random fields
on spins classified according to the ramp on which they turn
up is significantly modified from the uniform distribution given
in Equation (1). Indeed, the main object of this paper is to
calculate this distribution.

\section{Ramp-I}

Magnetization on ramp-I was determined earlier in reference 5
by exploiting a similarity between this problem and the problem
of random sequential adsorption(RSA)~\cite{evans}. The rate
equations of the RSA problem were used to determine $m(h_{a})$ on
ramp-I, but they could not determine $m(h_{a})$ on ramp-II and
ramp-III. Here we rederive the  result for ramp-I
by a different approach which can be extended to ramp-II
as well.

We start with a large negative applied field ($h_{a}=-\infty$)
when all spins are down and increase the applied field slowly.
At the start of ramp-I, i.e. at $h_{a}=-2|J|-\Delta$, the
spin with the largest positive quenched field becomes unstable
and flips up. As the applied field continues to increase, the
spin with the next largest quenched field turns up unless it
happens to be  next to a spin which is already up. In this
case the spin which turns up next is the one with the largest
quenched field from among the spins with both neighbors down.
Consider an arbitrary point on ramp-I at an applied
field $h_{a}=-2|J|-h$. In the following, the field h will be
used more frequently than $h_{a}$. In general, the field h
will be given by the relation  $h=-h_{a} \mbox{  mod  } |2J|$,
so that on each ramp it has the same range as the random field
($-\Delta \le h \le \Delta$). At an applied field $h_{a}=-2|J|-h$
all spins with quenched random field $h_{i} > h$ are relaxed,
and a fraction of them are up. The fraction of sites
with $h_{i} > h$ is given by
\be
p(h)=\int^{\Delta}_{h}p(h_{i})dh_{i} =\frac{\Delta-h}{2\Delta}
\ee

The fraction of spins which are up on ramp-I at
$h_{a}=-2|J|-h$ is given by
\be
P_\uparrow^{I}=\frac{1}{2}[1-e^{-2p}]
\ee

Let $P_{\downarrow\downarrow}^{I}$  be the probability (per site) of
finding a pair of adjacent down spins  on ramp-I at the applied
field $h_{a}=-2|J|-h$. It can be calculated as follows.
Imagine coloring all sites with $h_{i} > h$ black, and all
sites with $h_{i} < h$ white. Consider two adjacent down spins
A and B shown in Figure (2). The sites A and B can be both white,
both black, or mixed. Given that A is down, it is clear that the
state of B can only be influenced by the evolution of the system
to the right of B. Similarly, given that B is down, the state
of A can only be influenced by the evolution of the system to the
left of A. We shall refer to this as the principle of
conditional independence~\cite{mityushin}. It  requires
\be
P_{\downarrow\downarrow}^{I} =P(A\downarrow|B\downarrow)
P(B\downarrow|A\downarrow)
\ee
where $P(A\downarrow|B\downarrow)$ is the probability that
spin at site A is down given that spin at B is
down, and $P(B\downarrow|A\downarrow)$ is the probability that
B is down given that A is down. We take up the calculation of
$P(B\downarrow|A\downarrow)$. If B is a white site,
$P(B\downarrow|A\downarrow) = 1$ because the white sites have
not been relaxed from their initial state. If B is a black site and
the site to the right of B is a white site then
$P(B\downarrow|A\downarrow)=0$. In general
$P(B\downarrow|A\downarrow)$ depends on the length of the
string of black sites to the right of B. Suppose B is a
black site, and there are (n-1) additional black sites to
the right of B. In this case, the probability $P^{n}_{B}$
that B is down satisfies the following recursion relation,
\be
P^{n}_{B}=\frac{1}{n}P^{n-2}_{B}+(1-\frac{1}{n})P^{n-1}_{B}
\ee
The rationale for the above recursion relation is as follows.
Let the black site farthest from B on the right be labelled
as the n-th site. Any of the n sites could flip first. The
probability that the n-th site flips first is therefore
equal to $\frac{1}{n}$. If this happens, (n-1)-th site is
prevented from flipping up on ramp-I. The probability that B
is down is now reduced to the probability that the end point
of a chain of (n-2) black sites is down i.e. $P_{B}^{n-2}$.
This accounts for the first term in equation (8). The
probability that n-th site is not the first site to flip up
is equal to $(1-\frac{1}{n})$. Given this situation,
the probability that B is down is equal to the probability
that the end of a string of (n-1) black sites is down. This
accounts for the second term in equation (8). We can rewrite
the recursion relation (8) as
\be
(P^{n}_{B}-P^{n-1}_{B})=-\frac{1}{n}[P^{n-1}_{B}-P^{n-2}_{B}]
\ee

It has the solution,
\be
P^{n}_{B}=\sum^{n}_{m=0}\frac{(-1)^{m}}{m!}
\ee

Summing over various possible values of n with appropriate
weight, we get
\begin{eqnarray}
P_{B} & = & \sum^{\infty}_{n=0}\sum^{n}_{m=0}
\frac{(-1)^{m}}{m!}{p^{n}}(1-p) \nonumber \\
&  &   =\sum^{\infty}_{m=0}
\frac{(-1)^{m}}{m!}(1-p)\sum^{\infty}_{n=m}p^{n}
=\sum^{\infty}_{m=0}\frac{(-p)^{m}}{m!}
=e^{-p}
\end{eqnarray}
      
Thus,
\be
P_{\downarrow\downarrow}^{I}=e^{-2p}
\ee

Let $P_{\downarrow}^{I}$ be the probability per site of finding
a down spin and $P_{\downarrow\uparrow}^{I}$ the probability
per site of finding a down spin which is followed by an up spin.
Clearly,
\be
P_{\downarrow}^{I}=P_{\downarrow\downarrow}^{I}
+P_{\downarrow\uparrow}^{I}=1-P_{\uparrow}^{I}
\ee

Keeping in mind that on ramp-I an up spin must be preceeded
(as well as followed) by a down spin, we get
$P_{\downarrow\uparrow}^{I}=P_{\uparrow}^{I}$. Thus,
\bdm
P_\uparrow^{I}=1-P_{\downarrow\downarrow}^{I}-P_\uparrow^{I}
\edm
or,
\be
P_\uparrow^{I}=\frac{1}{2}[1-P_{\downarrow\downarrow}^{I}]
               =\frac{1}{2}[1-e^{-2p}]
\ee

The magnetization on ramp-I is given by
\be
m^{I}(h)=2P_\uparrow^{I}(h)-1=-e^{-2p(h)}
\ee

Equation (15) has been superposed on the simulation data
for ramp-I shown in Figure 1. The fit between the simulation
and theory is so good that the two curves are indistinguishable
from each other on the scale of Figure (1). The exact value of
the magnetization on plateau-I is equal to $-\frac{1}{e^{2}}$
which is approximately equal to -.135.
%%%%%%%%%%%%%%%%%%%%%%%%%%%%%%%%%%%%%%%%%%%%%%%%%%%%%
%%%%%%%%%%%%  Plateau-I  (platI.tex)  %%%%%%%%%%%%%%%
%%%%%%%%%%%%%%%%%%%%%%%%%%%%%%%%%%%%%%%%%%%%%%%%%%%%%

\section{Plateau-I}
Plateau-I contains down spins in singlets and doublets
punctuated by up spins. The down spins were relaxed on
ramp-I but did not turn up because the then applied
field was not strong enough to turn up spins with at
least one neighbor up. The singlets have two neighbors
up, and the applied field on ramp-II is still not
strong enough to turn them up. They have to wait for
their turn on ramp-III. On the other hand, each spin
in a doublet has one neighbor up and one down. It
therefore experiences zero net field from its neighbors.
The net field on a doublet spin is simply the sum of the
random field $h_{i}$ on its site and the applied
field $h_{a}=-h$. It turns up when $h_{a}+h_{i} \ge 0$, or
$h_{i} \ge h$. Note that
the random field is bound in the range
$-\Delta<{h_{i}}<+\Delta$. Therefore an applied field
smaller than $-\Delta$ is sufficiently negative to pin
down all doublets. This is the reason for the plateau
in the magnetization for applied fields in the
range $(-2|J|+\Delta)<h<{-\Delta}$. In each doublet,
the spin with the larger quenched field  $h_{i}$ flips
up on ramp-II when the applied field reaches a value
such that $h_{i} \ge h$. The spin with the smaller
quenched field then becomes a singlet which does not
flip up before ramp-III. Thus, in order to find the form of
ramp-II, we need to find the a posteriori distribution
of quenched random fields on the doublets.

Consider a doublet on plateau-I as shown in Figure (3).
The doublet sites are denoted as 1 and 2, and the
quenched random fields on these sites are
$h_{1}$ and $h_{2}$. The probability (per site) of finding
a doublet on plateau-I is easily obtained from equation (12)
by putting $p=1$. It is equal to $\frac{1}{e^{2}}$. We now
calculate the probability distribution for $h_{1}$ and $h_{2}$.
The distribution of $h_{1}$ and $h_{2}$ will be identical
if the evolution of the system on the two sides of the
doublet is similar to each other. We assume this to be
the case for now, although we shall examine a
more general situation later. In order to obtain the desired
probability distribution, consider the system on ramp-I
when all spins are relaxed upto an arbitrary applied field
$-2|J|-h$. At this point, the probability per site for
finding a doublet is $P_{\downarrow\downarrow}^{I}=e^{-2 p}$,
where p is fraction of sites with $h_{i}>h$ (black sites)
on an infinite lattice.
\bdm
p(h)=\frac{\Delta -h}{2\Delta}
\edm

It is often easier to think in terms of the
fraction p. We therefore introduce similar fractions
for the quenched fields given by the following relations.

\bdm
p(h_{i})=\frac{\Delta -h_{i}}{2\Delta}; \mbox{ (i=1,2,\ldots) } 
\edm

Given that site 1 is down on ramp-I, the conditional
probability that the adjacent site 2 is also down is
equal to $e^{-p}$. Site 2 may be black ($h_{2} > h$ or
equivalently $p_{2} < p$) with an a priori probability p,
or white with an a priori probability (1-p). If site 2 is
white $(h_{2} < h)$ , it must be down because white sites
are yet to be relaxed. Thus, the conditional probability
that the spin at site 2 is down, and the quenched
field at site 2 is larger than h is given by
\be
Prob(2\downarrow|1\downarrow,p_{2}< p)=\left[e^{-p}-(1-p)\right].
\ee

The probability that the quenched field $h_{2}$ lies in the
range h and h-dh, or equivalently  $p_{2}$ is in the range p and
p+dp can be obtained by taking the derivative of the above
expression. We obtain,
\be
Prob[p < p_{2} < p+dp] dp = [1-e^{-p}] dp
\ee
Similarly,
\be
Prob[p < p_{1} < p+dp] dp = [1-e^{-p}] dp
\ee

We now address an issue which is crucial for
determining ramp-II correctly. This concerns two adjacent
doublets on plateau-I as shown in Figure (4). Let $h_{1}$,
$h_{2}$, $h_{3}$, $h_{4}$ and $h_{5}$ denote the quenched
fields at sites 1,2,3,4,and 5 respectively. If $h_{2}>h_{1}$,
and $h_{4}>h_{5}$, then spins at sites 2, 3, and 4 will be
up at some value of the applied
field on ramp-II. When this happens, i.e when a triplet of
up spins is created on ramp-II, the central spin at site 3
becomes unstable and it flips down. It stays down till the
system reaches ramp-III. In order to take this effect into
account, we must know the probability per site of observing
two adjacent doublets on plateau-I, and also the distribution
of fields at sites 2 and 4.

A doublet on plateau-I has an important property. It
separates the lattice into two parts (one on each side of
the doublet) which have evolved uninfluenced by each other
on ramp-I. Thus, we can separate Figure (4) into three parts
as enclosed in the dashed boxes. Evolution inside each box has
remained shielded from the outside on ramp-I. The evolution
in the middle box requires that site 3 flips up before site 2
or site 4. The probability for this event is equal to
$\frac{1}{3}$. Given this event, the probability that spins
at sites 1 and 5 remain down all along ramp-I is each equal
to $\frac{1}{e}$. Thus the probability per site of observing
two adjacent doublets on plateau-I is
equal to $\frac{1}{3e^{2}}$. Note that it is quite
different from the square of the probability of finding
a single doublet!

The shielding property of the dashed boxes in
Figure (4) can also be used to calculate the a posteriori
distribution of fields $h_{1},h_{2},\ldots,h_{5}$. The
distribution of $h_{1}$ and $h_{5}$ is the same as obtained
above for a doublet with similar evolution on the two sides
(Figure 3).
\be
Prob(p \le p_{1} \le p+dp) dp = [1-e^{-p}] dp
\ee
and,
\be
Prob(p \le p_{5} \le p+dp) dp = [1-e^{-p}] dp
\ee

We now turn to the distributions of $h_{2}$,
and $h_{4}$. Suppose, in the middle box in Figure (4), the
central spin at site 3 is up when all sites in the system
with quenched fields $h_{i} \ge h$ (black sites) have been
relaxed. This necessarily means that $h_{3} \ge h$, i.e.
site 3 is a black site, but sites 2 and 4 have other options.
Site 3 will be up with probability $\frac{1}{3}p^{3}$ if
2 and 4 are both black, probability $p (1-p)^{2}$ if
2 and 4 are both white, and $(1-p) p^{2}$ if 2 and 4
are mixed. Thus the probability of observing two adjacent
doublets with the up spin separating them having field
$h_{3} \ge h$ ($p_{3} \le p) $ is given by

\be
P_{\downarrow\downarrow\uparrow\downarrow\downarrow}(p_{3} \le p)
= \left[ \frac{1}{3}p^{3}+p(1-p)^{2}+(1-p)p^{2}
\right] \frac{1}{e^{2}}
\ee

In order to calculate the distribution of $h_{2}$
and $h_{4}$, it is convenient to write the distributions
of the smaller and the larger of these two fields separately.
Without any loss of generality, we can assume
\bdm
h_{2}=min(h_{2},h_{4}) \mbox{   and    } h_{4}=max(h_{2},h_{4})
\edm
If $h_{2} \ge h$, then we have $h_{3} \ge h_{4} \ge h$ as well.
Thus the fraction of adjacent doublets on plateau-I with $h_{2}$,
$h_{3}$, and $h_{4}$ all greater than an arbitrary value h
is given by

\be
P_{\downarrow\downarrow\uparrow\downarrow\downarrow}
(p_{3} \le  p_{4} \le p_{2} \le p)
=  \frac{p^{3}}{3} \frac{1}{e^{2}} 
\ee

The above equation gives the cummulative fraction of $p_{2} \le p$
sites. The fraction of sites in the range $p+dp \ge p_{2} \ge p$
can be obtained by differentiating the above expression. We get,

\be
P_{\downarrow\downarrow\uparrow\downarrow\downarrow}
(p \le p_{2} \le p+dp)
= p^{2} \frac{1}{e^{2}} 
\ee

The distribution of $h_{4}$ is obtained similarly. We
find,

\be
P_{\downarrow\downarrow\uparrow\downarrow\downarrow}
(p \le p_{4} \le p+dp)
=2 p (1-p) \frac{1}{e^{2}}  
\ee

Given that we have a pair of adjacent doublets, the
probability densities of the quantities $p_{1}$,
$p_{2}$, $p_{4}$, and $p_{5}$ (each normalised to unity)
are given by:

\be
\rho_{1}(p_{1})=e \left[ 1-e^{-p_{1}} \right]
\ee

\be
\rho_{2}(p_{2})=3 p_{2}^{2}
\ee

\be
\rho_{4}(p_{4})=6 p_{4} (1-p_{4})
\ee

\be
\rho_{5}(p_{5})=e \left[ 1-e^{-p_{5}} \right]
\ee

%%%%%%%%%%%%%%%%%%%%%%%%%%%%%%%%%%%%%%%%%%%%%%%%%%%%%
%%%%%%%%%%%%  Plateau-I  (extra.tex)  %%%%%%%%%%%%%%%
%%%%%%%%%%%%%%%%%%%%%%%%%%%%%%%%%%%%%%%%%%%%%%%%%%%%%

In the following, we focus on adjacent doublets
which create up triplets on ramp-II. These are the
objects with $h_{2} \ge h_{1}$ and $h_{4} \ge h_{5}$.
At this stage, we can determine the lower and the upper
bound on these objects at any point on ramp-II when
spins with $h_{i} \ge h$ have been relaxed. Suppose we
order the adjacent
doublets in the order of increasing $h_{2}$ or
increasing $h_{4}$. Note that a sequence in increasing
$h_{2}$ does not posses any particular order in $h_{4}$,
and vice versa. The lower bound is given by the fraction
of objects with $h \ge h_{2} \ge h_{1}$ ($h_{4}$ being free to
have any value in the range $h_{2} \le h_{4} \le \Delta$).
We obtain
\bea
& & \int_{0}^{p}  \rho_{2}( p_{2} ) dp_{2}
\int_{p_{2}}^{1} \rho_{1}(p_{1}) dp_{1}  \nonumber \\
& & = (6 + 6 p + 3 p^{2}) e^{1-p}
+ ( 1+ e) p^{3} -\frac{3}{4} e p^{4} - 6e  \nonumber \\
& & =\left[ 16 - \frac{23}{4} e \right] \mbox{ (at $p=1$) }
=.369 \mbox { (approximately). }
\eea
Thus a minimum of approximately 37 \% adjacent doublets will
give rise to (unstable) up triplets on ramp-II.

Similarly, the fraction of adjacent doublets with
$h \ge h_{4} \ge h_{5} \ge -\Delta $  is given by
\bea
& & \int_{0}^{p}  \rho_{4}( p_{4} ) dp_{4}
\int_{p_{4}}^{1} \rho_{5}(p_{5}) dp_{5}  \nonumber \\
& & =(1+e) p^{2} (3 - 2 p) - 2 e p^{3} +\frac{3}{2} e p^{4}
+6 e -6 (1+p+p^{2}) e^{1-p} \nonumber \\
& & =\left[ \frac{13}{2} e - 17 \right] \mbox{ (at $p=1$) }
=.668 \mbox { (approximately). }
\eea
This gives the upper bound. No more than approximately
67 \% of the adjacent doublets can create (unstable)
up triplets on ramp-II.

%%%%%%%%%%%%%%%%%%%%%%%%%%%%%%%%%%%%%%%%%%%%%%%%%%%%%%%%%%%%
%%%%%%%%%%%%%%%%%%%%%%% rampII.tex %%%%%%%%%%%%%%%%%%%%%%%%%
%%%%%%%%%%%%%%%%%%%%%%%%%%%%%%%%%%%%%%%%%%%%%%%%%%%%%%%%%%%%
\section{Ramp-II}

Ramp-II is determined by the combination of two
opposite terms. The dominant term is the increase
in magnetization due to the decrease in the number of
doublets. When a doublet disappears, it adds an extra
up spin in the system which increases the magnetization.
Occasionally, a disappearing doublet creates
a string of three up spins. A triplet of up spins is
unstable on ramp-II, and the central spin of the triplet
flips down as soon as the triplet is created. This
decreases the magnetization. In the following, we
calculate the above two terms separately.

Refer to Figure (3) for calculating the first term.
The probability that the doublet shown in the Figure (3) 
disappears when spins with $h_{i} \ge h$ are relaxed
is given by
\be
P_{\uparrow\uparrow}^{II}=\frac{2}{e^2}
\int_{0}^{p}\rho(p_{1}) dp_{1}
\int_{p_{1}}^{1}\rho(p_{2}) dp_{2}
\ee
The factor $\frac{1}{e^2}$ is the probability per site of
finding a doublet shown in Figure (3). The factor 2 takes
care of the fact that either $h_{1}$ or $h_{2}$ may flip up
first. The integrals are written on the assumption that
the spin at site 1 flips up first. Together they give the
probability that $h_{1} \ge h$, and $h_{2} \le h_{1}$.
Note that when a doublet disappears, a pair of adjacent
up spins is created. This is the reason for the choice of
the subscript on $P_{\uparrow\uparrow}^{II}$. The
superscript indicates that the probability refers to
ramp-II. We obtain,

\be
P_{\uparrow\uparrow}^{II}= \frac{1}{e^2}
- \left[\left(1+e^{-1}\right) -                   
\left(p+e^{-p}\right) \right]^{2}
\ee

We now caculate the fraction of (unstable) up triplets
formed in the system when all spins with $h_{i} \ge h$
have been relaxed on ramp-II. Let us refer to Figure (4).
Recall that $h_{2} \le h_{4}$ in this figure. We want the
probability that $h_{2} \ge h_{1}$, $h_{4} \ge h_{5}$,
and $h_{2} \ge h$. This is given by,

\be
P_{\uparrow\uparrow\uparrow}^{II}=\frac{1}{3 e^2}
\int_{0}^{p}\rho(p_{2}) dp_{2}
\int_{p_{2}}^{1}\rho(p_{1}) dp_{1}
\left [
\frac
{\int_{0}^{p_{2}}\tilde\rho(p_{4}) dp_{4}
\int_{p_{4}}^{1}\rho(p_{5}) dp_{5}}
{\int_{0}^{p_{2}}\tilde\rho(p_{4}) dp_{4} }
\right ]
\ee
The first factor is the probability per site of finding the
object shown in Figure (4). The next two integrals give the
probability that $h_{2} \ge h$, and $h_{1} \le h_{2}$. The
quantity in the square brackets is understood as follows:
When $h_{2}$ is in the range h and h+dh, $h_{4}$ can be
anywhere in the range $h_{2}$ to $\Delta$. Let $\tilde\rho(h_{4})$
be the density of $h_{4}$ in this range. Clearly,
\be
\rho_{2}(h_{2})=\int_{h_{2}}^{\Delta} \tilde\rho_{4}(h_{4})
\frac{dh_{4}}{2\Delta}
\ee
or,
\be
\rho_{2}(p_{2})=\int_{0}^{p_{2}} \tilde\rho_{4}(p_{4}) dp_{4}
\ee
Thus,
\be
\tilde\rho_{4}(p_{2})=\frac{d\rho_{2}(p_{2})}{ dp_{2}}
\ee
or,
\be
\tilde\rho_{4}(p_{4})=6 p_{4}
\ee
We get,
\bdm
P_{\uparrow\uparrow\uparrow}^{II} =\frac{1}{3}
\left[
        \frac{3}{2} + \frac{6}{e}
        - 6 \left( 1 + \frac{1}{e} \right) p
        + 3 p^{2}
        + {\left( 1 + \frac{1}{e} \right)}^{2} p^{3}
        -\frac{5}{4} \left( 1 + \frac{1}{e} \right) p^{4}
        +\frac{2}{5} p^{5} \right.  \\
\edm
\bdm
         \left. -\left\{ 6 \left( 1 + \frac{1}{e} \right)
        -6 p -3 \left( 1 + \frac{1}{e} \right) p^{2}
        + 2 p^{3} \right\} e^{-p}
        +\left(\frac{9}{2} + 3 p\right) e^{-2p}
\right]
\edm
We show in Figure (5) a comparison of the above expression
with a result from the simulation. As may be expected, the
agreement between the simulation and the theory is excellent.

Incidentally, an interesting quantity is the ratio of
(unstable) up triplets at the end of ramp-II to the
fraction of adjacent doublets at the start of
ramp-II. We noted in the previous section that this
ratio must lie approximately in the range .369 to .668.
If there were no correlations between adjacent
doublets, this ratio would be equal to $\frac{1}{4}$,
because the events $h_{2} > h_{1}$, or $h_{4} > h_{5}$
would occur with probability $\frac{1}{2}$. The exact
value of the probability that $h_{4} \ge h_{5} \ge -\Delta$
and $h_{2} \ge h_{1} \ge -\Delta$ is given by

\be
P_{\uparrow\uparrow\uparrow}^{II}
=\frac{1}{3 e^{2}} \left[ \frac{11}{2}
+ \frac{7}{4} e -\frac{27}{20} e^{2}
\right]
\ee
The quantity in the square brackets is approximately
equal to .281. We have checked this result rather carefully
numerically, and it is born out by the simulations.
Finally, putting the various terms together, the
probability of an up spin on ramp-II is given by
\be
P_{\uparrow}^{II}(p) = P_{\uparrow}^{I}(1) +
P_{\uparrow\uparrow}^{II}(p) -
P_{\uparrow\uparrow\uparrow}^{II}(p)
\ee
The magnetization on ramp-II is given by
\be
m^{II}(p)=2 P_{\uparrow}^{II}(p) -1
\ee

This expression has been superposed on the numerical
data for ramp-II shown in Figure (1). The agreement
between the numerical data and the theory is
extremely good. The exact value of the magnetization on
plateau-II, and its numerical estimate are given by
\be
m^{II}(1)= \left[ \frac{27}{30} -\frac{7}{6} e^{-1}
-\frac{8}{3} e^{-2} \right] = .109
\mbox{  (approximately) }
\ee

%%%%%%%%%%%%%%%%%%%%%%%%%%%%%%%%%%%%%%%%%%%%%%%%%%%%%%%
%%%%%%%%%%%%  Plateau-II  (platII.tex)  %%%%%%%%%%%%%%%
%%%%%%%%%%%%%%%%%%%%%%%%%%%%%%%%%%%%%%%%%%%%%%%%%%%%%%%

\section{Plateau-II}
Each down spin on plateau-II is a singlet. However,
there are three different classes of singlets:
the singlets formed on ramp-I; singlets formed
on ramp-II by a vanishing doublet; and finally the
singlets formed on ramp-II by the
unstable central spin of an up triplet flipping down.
Each class is characterized by its own a posteriori
distribution of the random field.

Let us denote the three distribution densities by
$\rho^{II}_{1}$, $\rho^{II}_{2}$, and $\rho^{II}_{3}$
respectively. Let $P^{II}_{1}$, $P^{II}_{2}$, and
$P^{II}_{3}$ denote the cummulative populations in each
class when spins with $h_{i} \ge h$ have been relaxed
on ramp-III. We have,

\be
P^{II}_{i}(p)=\int_{0}^{p}\rho^{II}_{i}(p_{i}) dp_{i},
\mbox{   i=1,2,3. }
\ee

It is useful to think of the singlets in each class as
being black $(h_{i} \ge h)$, or white $(h_{i} < h)$,
where $h_{i}$ is the quenched random field at the
singlet site, and h is an arbitrary reference field.
In order to calculate ramp-III, we need only the
populations of black singlets in each class given by
$P^{II}_{i}(p)$. If needed, one can obtain the density
$\rho^{II}_{i}(p)$ by differentiating $P^{II}_{i}(p)$
with respect to p.

The fraction of black singlets created on ramp-I is
given by,
\be
P^{II}_{1}(p)=p-\frac{1}{2}\left[1-e^{-2p}\right]
-\frac{2}{e}\left[e^{-p}-(1-p)\right]
\ee
The explanation of the above equation is as follows.
Imagine ordering the sites of the lattice in order of
decreasing quenched field on the site. When all sites
with $h_{i} \ge h$ have been relaxed, the fraction of
the relaxed sites is equal to p (the black sites).
This fraction is made of the up sites (the second
term on the right), black doublet sites (the last term),
and the black singlets. Hence the equation for
$P^{II}_{1}(p)$. The last term
is written as follows. In each doublet, there are two
sites from which we can choose one. This accounts for
the factor 2. The quantity in the square bracket gives
the probability that the chosen site is black, and
$\frac{1}{e}$ is the probability that the other site
can have any allowed value of the quenched field.

The fraction of black singlets generated by vanishing
doublets on ramp-II is given by,
\be
P^{II}_{2}(p)=\left[e^{-p}-(1-p)\right]^2
\ee
The above equation is easily understood. It is the
probability that both sites of the doublet are black.
If both sites of the doublet are black, the one with
higher random field must flip up on ramp-II, leaving
us with a singlet on plateau-II which is black.

The fraction of black singlets created by unstable
triplets requires the calculation of triplets. We
have calculated the fraction of triplets as
they are formed on ramp-II. What we need now is a
similar but different calculation. The point can be
understood with a reference to Figure (4). Recall
that $h_{3} \ge h_{4} \ge h_{2}$. On ramp-II, we
needed the fraction of triplets with $h_{2} \ge h$,
because the formation of triplets is controlled by
this threshold. The restoration of the triplets on
ramp-III is  controlled by the condition
$h_{3} \ge h$. In the earlier calculation, sites 2, 3,
and 4 were all black sites. In the calculation
needed now, only site 3 is black. Sites 2, and 4 are
white, and we want $h_{1} \le h_{2}$, and
$h_{5} \le h_{4}$. The probability for this event is
given by

\bdm
P^{II}_{3}(p)=2 \int_{0}^{p}dp_{3}
\int_{p_3}^{1}  dp_{4}
\int_{p_4}^{1} \left[ 1-e^{-p_{5}} \right] dp_{5}
\int_{p_4}^{1} dp_{2}
\int_{p_2}^{1} \left[ 1-e^{-p_{1}} \right] dp_{1}
\edm

As a check, we note that
\bdm
2 \int_{p_3}^{1}  dp_{4}
\int_{p_4}^{1} dp_{2} = (1-p_{3})^{2}
\edm
Thus, the double integral gives the probability that
sites 2 and 4 are white ( compare with the second term on the
right hand side of Equation 21 ). The extra terms in
$P^{II}_{3}(p)$ take into account the requirements
$h_{5} \le h_{4}$, and $h_{1} \le h_{2}$.

\bdm
P^{II}_{3}(p)=2 \int_{0}^{p}dp_{3}
\int_{p_3}^{1} \left[ 1+e^{-1}
- p_{4}-e^{-p_{4}} \right] dp_{4}
\int_{p_4}^{1} \left[ 1+e^{-1}
- p_2-e^{-p_{2}} \right] dp_2
\edm
We get,

\bea
P^{II}_{3}(p)= & -\left( 1 + \frac{2}{e} \right)
+ \left[ \frac{1}{4} + \frac{2}{e}
+ \frac{4}{e^{2}} \right] p
- \frac{1}{2} \left[ 1 + \frac{5}{e}
+ \frac{4}{e^{2}} \right] p^2
+ \frac{1}{3} \left[ \frac{3}{2}
+ \frac{4}{e} + \frac{1}{e^{2}}
\right] p^3 \nonumber \\ &
-\frac{1}{4} \left[ 1+\frac{1}{e} \right] p^4
+ \frac{1}{20} p^5 + \left[1+\frac{2}{e}\right]e^{-p}
-\frac{2}{e}p e^{-p} +p^2 e^{-p}
+\frac{1}{2} \left[ 1-e^{-2p} \right]
\eea

As a check we note that,
\bdm
P^{II}_{3}(1)= \frac{1}{3e^2} \left[
\frac{11}{2} +\frac{7}{4} e -\frac{27}{20} e^2 \right]
\edm
This is the same as given by Equation (38). As one may
expect, the number of triplets with $h_{3} \ge h$ is larger 
than triplets with $h_{2} \ge h$, but the two merge
at $h = -\Delta$ i.e. $p=1$.

It is of some interest to calculate the number of triplets
with $h_{4} \ge h$ as well, although this quantity is not
directly needed in the calculation of ramp-III. Denoting
this quantity by $P^{II}_{4}(p)$, we get

\bdm
P^{II}_{4}(p)=2 \int_{0}^{p}p_{4}dp_{4}
\int_{p_4}^{1} \left[ 1-e^{- p_{5}} \right] dp_{5}
\int_{p_4}^{1} dp_{2} \int_{p_{2}}^{1}
\left[ 1 -e^{-p_{1}} \right]dp_{1}
\edm

Note that either $h_{2}$ or $h_{4}$ could have been the
larger of the two fields although we have assumed that
$h_{4}$ is the larger one. This accounts for the factor 2
on the right hand side. The factor $p_{4}$ comes from
integrating $p_{3}$ over the range $0$ to $p_{4}$.
The other terms are self-explanatory. We get,

\bea
P^{II}_{4}(p)= & -\left[ \frac{1}{2} + \frac{2}{e} \right]
+ \frac{1}{2} \left[ 1 + \frac{5}{e} + \frac{4}{e^{2}} \right]p^{2}
- \frac{2}{3} \left[ \frac{3}{2} + \frac{4}{e}
+ \frac{1}{e^{2}} \right]p^{3}
+ \frac{3}{4} \left[ 1 + \frac{1}{e} \right]p^{4} \nonumber \\
& -\frac{1}{5}p^{5}
+\left[ \left( 1 + \frac{2}{e} \right) \left( 1 +p +p^2 \right)
+ p^{3} \right] e^{-p}
-\left[ \frac{1}{2} + p \right] e^{-2p}
\eea
It can be checked that $P^{II}_{4}(1)$ also reduces to the
expression in Equation (38) as may be expected. In Figure (5), we
have shown the theoretical expressions for
$P^{II}_{\downarrow\downarrow\downarrow}(p)$,  $P^{II}_{3}(p)$,
and $P^{II}_{4}(p)$ along with the results from the simulations
for the same quantities. The simulations were performed for
a system of $10^{3}$ spins, and averaged over $5 \cdot 10^{3}$
different realizations of the random field distribution. The
agreement is excellent as may be expected from an exact analytic
result. The agreement between the simulation and the theory also
justifies (albeit post facto) the implicit assumption in our
analysis that the system is self-averaging. The fact that
simulations over a relatively small size of the system
($10^{3}$ spins) agree with the exact result is due to the
super-exponential decay of correlations in this system~\cite{evans}.

%%%%%%%%%%%%%%%%%%%%%%%%%%%%%%%%%%%%%%%%%%%%%%%%%%%%%%%%%%%%
%%%%%%%%%%%%%%%%%%%%%%% rampIII.tex %%%%%%%%%%%%%%%%%%%%%%%%
%%%%%%%%%%%%%%%%%%%%%%%%%%%%%%%%%%%%%%%%%%%%%%%%%%%%%%%%%%%%
\section{Ramp-III}

We are not in a position to write the analytic expression for
ramp-III at this stage. In the previous section, we have
obtained some ingredients which are necessary for this
purpose, but these are not sufficient. If the only process
occurring on ramp-III were the gradual decrease in the number
of singlets initially present on plateau-II, then we have the
required information to determine ramp-III. However, additional
singlets are created (and later destroyed) on ramp-III. Let us
illustrate this with an example. Consider a down spin on
plateau-II, say with random field $h_{d}$. Its nearest neighbors
are necessarily up because plateau-II contains only singlets.
The next nearest neighbors can be up or down. Consider a nearest
neighbor, and let $h_{u}$ be the random field on it.
The site with field $h_{u}$ may have one or both of its nearest
neighbors down. Suppose one of the nearest neighbors of $h_{u}$
is up. Now consider the situation when the singlet just flips up
on ramp-III at an applied field $h_{a}$, i.e.
$-2 |J| +h_{d} +h_{a} = \epsilon$, where $\epsilon \ge 0$.
The applied field at this point is
$h_{a} =2 |J| - h_{d} + \epsilon$. If $h_{u}$ is to remain up
after $h_{d}$ has flipped up, we must have
$h_{u} - 2 |J| + h_{a} \ge 0$, or $h_{u} -h_{d} + \epsilon \ge 0$.
This is not guaranteed if $h_{u} \le h_{d}$. In this
case $h_{u}$ will flip down when $h_{d}$ flips up, and later flip up
again in a higher applied field. The conditional probability distributions
which govern the creation of these new singlets and their restoration
on ramp-III require further careful investigation. We hope to take up
this work in a future article.

%%%%%%%%%%%%%%%%%%%%%%%%%%%%%%%%%%%%%%%%%%%%%%%%%%%%%%%%%%%%
%%%%%%%%%%%%%%%%%%%%%%% conclu.tex %%%%%%%%%%%%%%%%%%%%%%%%%
%%%%%%%%%%%%%%%%%%%%%%%%%%%%%%%%%%%%%%%%%%%%%%%%%%%%%%%%%%%%
\section{Conclusion}
We have presented an exact solution of the zero-temperature
dynamics of the one-dimensional anti-ferromagnetic random
field Ising model in a slowly increasing applied field for
a limited range of the applied field. This
problem was posed a few years ago, but to the best of our
knowledge remained unsolved in this period. The solution presented
here complements the analytic solution of the non-equilibrium
dynamics of the ferromagnetic random field Ising model.
The key difference from the ferromagnetic case is
the absence of avalanches. A large number of physical systems
relax by avalanches, but clearly not all of them. Therefore,
models of relaxation without avalanches have to be studied as
well. The analysis presented here is a small step in this
direction. It also attempts to develop the techniques used in
the study of the problem of random sequential adsorption (RSA).
The relationship between ramp-I and RSA was noted earlier.
Indeed, the RSA rate equations were used to solve the
problem of ramp-I. However, the problem of ramp-II and
ramp-III could not be solved. Here we have solved
the problem of ramp-I by an alternate (although not
profoundly different) method which focuses on the
a posteriori modification of the random field distribution
by the dynamical selection process. This method yields
more detailed information on ramp-I than was available
earlier, and allows us to get an exact expression for ramp-II.
We hope that the method employed here will enable us to solve
the problem of ramp-III in the near future, and it may be
useful in other related problems as well.

We thank D Dhar for useful discussions during a visit to NEHU.

%%\bibliographystyle{unsrt}

%%%%%%%%%%%%% fcaps.tex  (Figure Captions)
\newpage
\Large{Figure Captions}

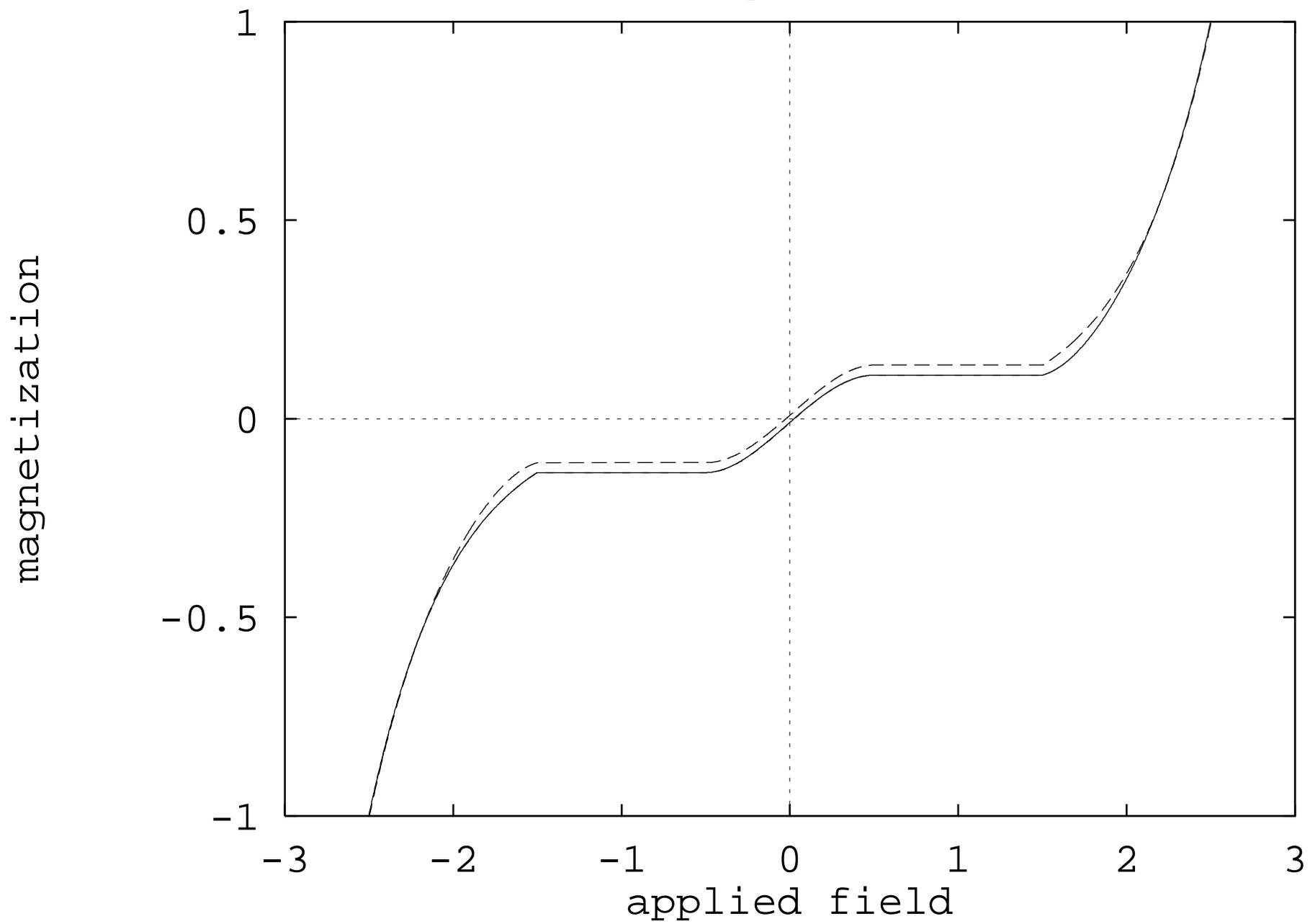
Figure 1: \\ Magnetization $m(h_{a})$ in an applied
field $h_{a}$  for an anti-ferromagnetic RFIM ($J=-1$),
obtained from a simulation of $10^{3}$ spins
averaged over $10^{3}$ different realizations of the random
field distribution of width $2\Delta=1$. The solid line
shows the magnetization in increasing field. The broken
line shows the return half of the hysteresis loop. The
data in increasing field is separated into five parts along
the applied field:
Ramp-I ($h_{a}=-2|J|-\Delta$ to $h_{a}=-2|J|+\Delta$);
Plateau-I ($h_{a}=-2|J|+\Delta$ to $h_{a}=-\Delta$);
Ramp-II ($h_{a}=-\Delta$ to $h_{a}=+\Delta$);
Plateau-II ($h_{a}=+\Delta$ to $h_{a} = 2|J|-\Delta$);
and Ramp-III ($h_{a}=2|J|-\Delta$ to $h_{a}=2|J|+\Delta$).
Theoretical expressions obtained in the paper have been
superposed on Ramp-I, Plateau-I, Ramp-II, and Plateau-II
in increasing field.

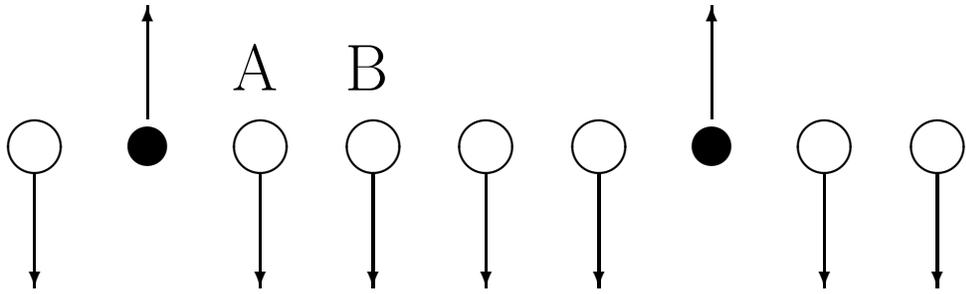
Figure 2: \\ Spins on Ramp-I in an applied field $-2|J|-h$.
Filled circles show sites with quenched field $h_{i} > h$.
The probability per site of a doublet (two adjacent down spins)
such as AB is equal to $e^{-2p}$, where p is the fraction of
filled circles on the infinite lattice.

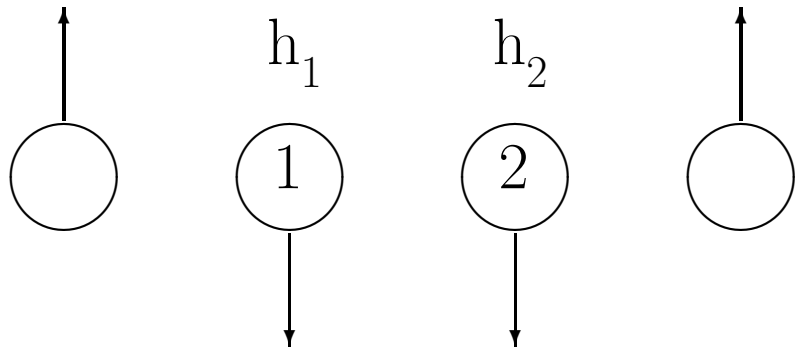
Figure 3: \\ A doublet on Plateau-I: $h_{1}$ and $h_{2}$ are the
quenched random fields on the doublet sites 1 and 2 respectively.

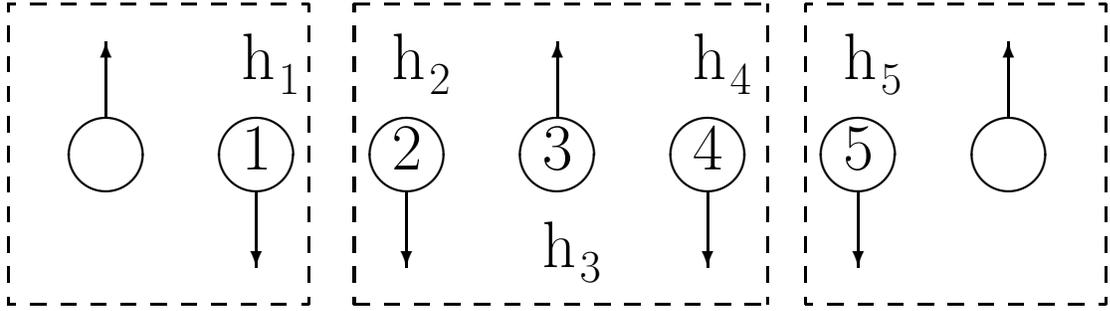
Figure 4: \\ Two adjacent doublets on Plateau-I: Each doublet
separates the lattice into two parts whose evolution histories
on Ramp-I are independent of each other. Evolutions inside
each dashed box is shielded from outside. The probability that
spin at site 3 flips up on Ramp-I is therefore equal to
$\frac{1}{3}$. Given this, the probability that the spins at
sites 1 and 5 remain down all along Ramp-I is equal to
$\frac{1}{e}$ each. The shielding property of the boxes can also
be used to determine a posteriori distribution of random fields
$h_{1}, h_{2}, h_{3}, h_{4}, \mbox{  and  } h_{5}$.

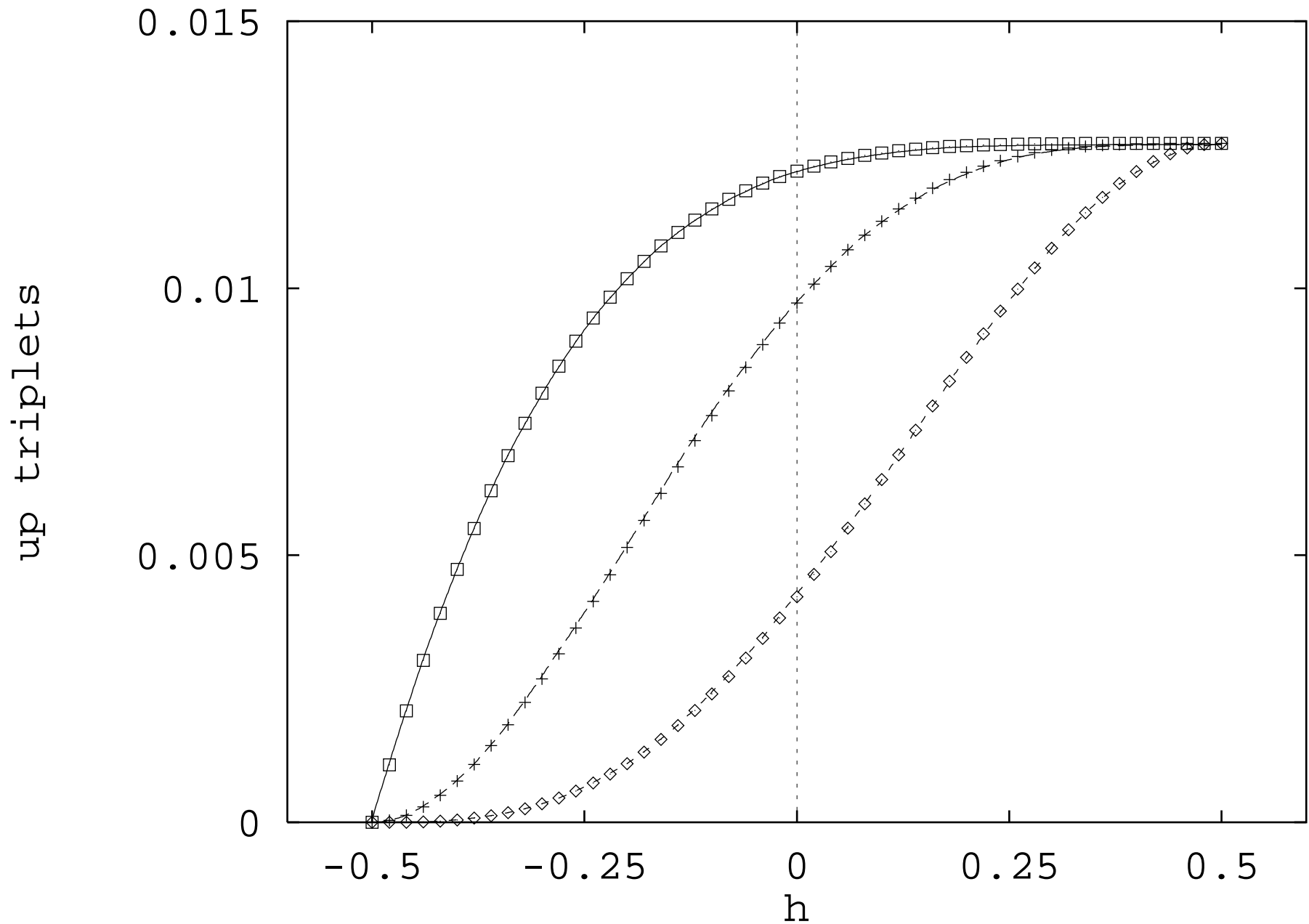
Figure 5: \\ Unstable up triplets on Ramp-II with
$h_{2} \ge h$ (lower curve);
$h_{4} \ge h$ (middle curve); and
$h_{3} \ge h$ (upper curve). Refer to Figure 4 for
$h_{2}$, $h_{4}$, and $h_{3}$. The simulation data is
shown by lines, and the symbols on each line show the
corresponding theoretical prediction.

}

%%%%%\end{document}
%%%%%%%%%%%%%%%%%%%%%%%%%%%%%%%%%%%%%%%%%%%%%%%%%%%%%%%%
%%%%%%%%%%%%%%%%%%%%    Figure  2   %%%%%%%%%%%%%%%%%%%%
%%%%%%%%%%%%%%%%%%%%%%%%%%%%%%%%%%%%%%%%%%%%%%%%%%%%%%%%
%%%\documentstyle{article}
%%%\begin{document}
\newpage
\thispagestyle{empty}
\setlength{\unitlength}{1.5cm}
\begin{picture}(15.,2.)
\thicklines
\put(0,1){\circle{.5}} \put(0,.75){\vector(0,-1){1}}
\put(1,1){\circle*{.5}} \put(1,1.25){\vector(0,1){1}}
\put(2,1){\circle{.5}} \put(2,.75){\vector(0,-1){1}}
\put(3,1){\circle{.5}} \put(3,.75){\vector(0,-1){1}}
\put(4,1){\circle{.5}} \put(4,.75){\vector(0,-1){1}}
\put(5,1){\circle{.5}} \put(5,.75){\vector(0,-1){1}}
\put(6,1){\circle*{.5}} \put(6,1.25){\vector(0,1){1}}
\put(7,1){\circle{.5}} \put(7,.75){\vector(0,-1){1}}
\put(8,1){\circle{.5}} \put(8,.75){\vector(0,-1){1}}
\put(1.75,1.5){\huge{A}}
\put(2.75,1.5){\huge{B}}
\put(3,-2) {\bf{\huge{Figure 2}}}
\end{picture}
\vspace{2cm}
%%%%%%%%%%%%%%%%%%%%%%%%%%%%%%%%%%%%%%%%%%%%%%%%%%%
%%%%%%%%%%%%%%%%%  figure 3  %%%%%%%%%%%%%%%%%%%%%%
%%%%%%%%%%%%%%%%%%%%%%%%%%%%%%%%%%%%%%%%%%%%%%%%%%%
\begin{picture}(15.,6.)
\thicklines
\put(1,-2){\circle{1}}
\put(.85,-2.1){\huge{}} \put(1,-1.5){\vector(0,1){1}}
\put(3,-2){\circle{1}}
\put(2.85,-2.1){\huge{1}} \put(3,-2.5){\vector(0,-1){1}}
\put(5,-2){\circle{1}}
\put(4.85,-2.1){\huge{2}} \put(5,-2.5){\vector(0,-1){1}}
\put(7,-2){\circle{1}}
\put(6.85,-2.1){\huge{}} \put(7,-1.5){\vector(0,1){1}}
\put(2.8,-1){\huge{h}}
\put(3.1,-1.2){\bf{\Large{1}}}
\put(4.8,-1){\huge{h}}
\put(5.1,-1.2){\bf{\Large{2}}}
\put(3,-5) {\bf{\huge{Figure 3}}}
\end{picture}
%%%\end{document}
%%%%%%%%%%%%%%%%%%%%%%%%%%%%%%%%%%%%%%%%%%%%%%%%%%%%%%%%
%%%%%%%%%%%%%%%%%%%%%%%% Figure 4 %%%%%%%%%%%%%%%%%%%%%%
%%%%%%%%%%%%%%%%%%%%%%%%%%%%%%%%%%%%%%%%%%%%%%%%%%%%%%%%
%%%\documentstyle[12pt]{article}
%%%\begin{document}
\newpage
\thispagestyle{empty}
\setlength{\unitlength}{1cm}
\begin{picture}(15,4)
\thicklines
\put(1,1){\circle{1}}
\put(3,1){\circle{1}}
\put(5,1){\circle{1}}
\put(7,1){\circle{1}}
\put(9,1){\circle{1}}
\put(11,1){\circle{1}}
\put(13,1){\circle{1}}
\put(1,1.5){\vector(0,1){1}}
\put(3,.5){\vector(0,-1){1}}
\put(5,.5){\vector(0,-1){1}}
\put(7,1.5){\vector(0,1){1}}
\put(9,.5){\vector(0,-1){1}}
\put(11,.5){\vector(0,-1){1}}
\put(13,1.5){\vector(0,1){1}}
\put(-.3,-1){\dashbox{0.2}(4,4)}
\put(4.3,-1){\dashbox{0.2}(5.5,4)}
\put(10.3,-1){\dashbox{0.2}(4,4)}
\put(2.8,.8){\huge{1}}
\put(4.8,.8){\huge{2}}
\put(6.8,.8){\huge{3}}
\put(8.8,.8){\huge{4}}
\put(10.8,.8){\huge{5}}
\put(2.8,2){\huge{h}}
\put(3.3,1.8){\bf{\Large{1}}}
\put(4.8,2){\huge{h}}
\put(5.3,1.8){\bf{\Large{2}}}
\put(6.8,-.5){\huge{h}}
\put(7.3,-.7){\bf{\Large{3}}}
\put(8.8,2){\huge{h}}
\put(9.3,1.8){\bf{\Large{4}}}
\put(10.8,2){\huge{h}}
\put(11.3,1.8){\bf{\Large{5}}}
\put(5.5,-6) {\bf{\huge{Figure 4}}}
\end{picture}


\begin{thebibliography}{99}

\bibitem{sethna}J P Sethna, K Dahmen, S Kartha, J A Krumhansl,
B W Roberts, and J D Shore, Phys Rev Lett 70, 3347 (1993).

\bibitem{dhar} D Dhar, P Shukla, and J P Sethna, J Phys A:
Math. Gen. 30,5259 (1997).

\bibitem{shukla1} P Shukla, Physica A 233, 235 (1996).

\bibitem{kisker} J Kisker,H Rieger, and H Schreckenberg, J Phys A:
Math. Gen. 27, L853 (1994). This paper discusses the non-equilibrium
dynamics of a non-random, one-dimensional Ising model, with
three-spin interactions, at low temperatures. It is qualitatively
similar to the dynamics of the random field Ising model at
zero-temperature.

\bibitem{shukla2} P Shukla, Physica A 233, 242 (1996).

\bibitem{sanjib} S Sabhapandit, P Shukla, and D Dhar, submitted for
publication.

\bibitem{evans} J W Evans, Rev Mod Phys 65, 1281 (1993).

\bibitem{mityushin} L G Mityushin, Prob Peredachi Inf 9, 81 (1973).
Also see reference (7) for a detailed discussion of the screening
property of this class of problems.
\end{thebibliography}
\end{document}